\documentclass[conference]{IEEEtran}
\usepackage{graphicx}
\usepackage{amsmath,amssymb,dsfont,bbm,epstopdf,pgfplots,mathtools,enumitem,mathrsfs, bbm, algpseudocode , graphicx, dblfloatfix, booktabs, physics, algorithm,algcompatible, derivative, physics} 
\usepackage[normalem]{ulem}
\usepackage{color}
\usepackage{cite}
\usepackage{graphics}
\usepackage{tikz}
\usepackage{pgfplots}
\usepackage{subcaption}
\usepackage{bbm}
\usepackage[utf8]{inputenc} 
\usepackage[T1]{fontenc}
\usepackage[left=0.6in,right=0.6in,top=0.71in, bottom = 0.94in]{geometry}
\usetikzlibrary{shapes, arrows, decorations.markings, arrows.meta}
\usetikzlibrary{patterns} 
\allowdisplaybreaks
\graphicspath{{.}{./Figures/}} 
\allowdisplaybreaks[4] 

\newtheorem{theorem}{Theorem}
\newtheorem{proposition}{Proposition}

\newtheorem{definition}{Definition}

\newcommand{\R}{{\mathbb R}}

\renewcommand{\Pr}{{\mathbb{P}}}

\definecolor{ForestGreen}{rgb}{0.0, 0.5, 0.0}

\newcommand{\D}{\mathsf{D}}
\newcommand{\C}{\mathsf{C}}
\renewcommand{\R}{\mathsf{R}}

\begin{document}
\title{ A Memory-Based Reinforcement Learning Approach to Integrated Sensing and Communication}
\author{\IEEEauthorblockN{Homa Nikbakht$^{1}$, Mich\`ele Wigger$^2$, Shlomo Shamai (Shitz)$^3$, and H.~Vincent Poor$^1$}
	\IEEEauthorblockA{$^1$Princeton University,   $\quad ^2$LTCI,   {T}$\acute{\mbox{e}}$l$\acute{\mbox{e}}$com Paris, IP Paris,  $\quad ^3$Technion \\
		\{homa, poor\}@princeton.edu, michele.wigger@telecom-paris.fr,  sshlomo@ee.technion.ac.il}}
\maketitle

 \begin{abstract}
In this paper, we consider a point-to-point integrated sensing and communication (ISAC) system, where a transmitter conveys  a message to a receiver over a channel with memory and simultaneously estimates the state of the channel through the backscattered signals from the emitted waveform. Using Massey’s concept of directed information for channels with memory, we formulate the capacity-distortion tradeoff for the ISAC problem when sensing is performed in an online fashion.  Optimizing the transmit waveform for this system to simultaneously achieve good communication  and  sensing performance is a complicated task, and thus we propose a deep reinforcement learning (RL) approach to find a solution. The proposed approach enables the agent to optimize the ISAC performance by learning a reward that reflects the difference between the communication gain and the sensing loss. Since the state-space in our RL model is \`a priori unbounded, we employ  deep deterministic policy gradient  algorithm (DDPG). Our numerical results suggest a significant performance improvement when one considers  unbounded state-space as opposed to a simpler RL problem with reduced state-space. In the extreme case of degenerate state-space only memoryless signaling strategies are possible. Our results thus emphasize the   necessity of well exploiting the memory inherent in  ISAC systems.
\end{abstract}
\section{Introduction}
Integrating sensing and communication (ISAC) into a single system is motivated by reducing hardware costs, bandwidth usage, and power consumption. It is enabled by several features anticipated for 6G communication systems: higher frequency bands (from mmWave up to THz), wider bandwidths and denser distributions of massive antenna arrays\cite{Visa2024,Luo2024, Liu2022, Ahmadi2024,Calderbank2023}. 

Recently, deep learning technology has demonstrated its capability in various wireless communication applications such as channel estimation, signal detection, and resource allocation \cite{Poorbook, Respati2024}. Motivated by this, some recent studies have focused on enhancing the performance of an ISAC system using deep reinforcement learning approaches in different model-based and model-free settings  and for a wide range of applications \cite{Pulkkinen2024,PulkkinenICASSP,Nguyen2024}.

 In this paper, we propose a reinforcement learning (RL) approach to study fundamental limits of  ISAC systems with memory by adopting a deep deterministic policy gradient (DDPG) algorithm \cite{Lillicrap2015} where an agent simultaneously takes sensing and communication actions to optimize the ISAC performance. More specifically, we  use Massey’s concept of directed information for channels with memory \cite{Massey1990} and formulate the capacity-distortion trade-off under an online-sensing framework. This formulation includes an optimization problem where the objective is to optimize the transmit waveform so as to simultaneously achieve good communication and sensing  performance. We simplify this optimization problem for the class of \emph{unifilar channels} \cite{Gallager1968}, and reformulate the  problem in this special case as a Markov decision process  (MDP). Finally, using a DDPG algorithm, we numerically evaluate the capacity-distortion trade-off for a specific example. 
 Our numerical results suggest a significant performance improvement if our RL approach is applied to the full model with unbounded state-space as opposed to a restricted and simplified model with limited  state-space. In the extreme case of degenerate state-space, the RL formulation only allows for memoryless signaling strategies and is highly suboptimal.
 
The capacity-distortion trade-off of  ISAC systems has been studied both in the asymptotic \cite{Kobayashi2019,Ahmadipour2022, Chen2023} and finite blocklength \cite{Nikbakht2024} regimes. However, while most works focus on  memoryless ISAC channels, \cite{Chen2023} considered a very general model with memory similar to the one in this work.   The difference between our work and \cite{Chen2023} lies in our focus on online-estimators that sense the targets in an online manner and not just at the end of the communication. Moreover, we managed to simplify the general complicated expression for the class of unifilar channels. 

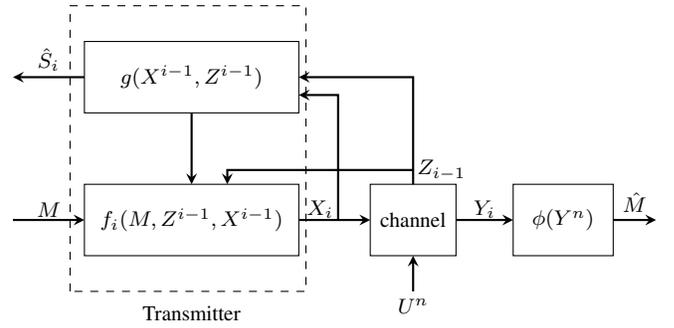
\begin{figure}[t]
  \centering
\begin{tikzpicture}[scale=0.95, >=stealth]
\centering
\footnotesize
\tikzstyle{every node}=[draw,shape=circle, node distance=0.5cm];
\draw [thick, ->] (0,0.5)--(1,0.5);
\node[draw =none] at (0.5,0.65) {$M$};
\draw (1,0) rectangle (4,1);
\node[draw =none] at (2.5,0.5) {$f_i(M, Z^{i-1}, X^{i-1})$};
\draw [thick, ->] (4,0.5)--(5,0.5);
\node[draw =none] at (4.3,0.65) {$X_i$};
\draw (5,0) rectangle (6.2,1);
\node[draw =none] at (5.6,0.5) {\footnotesize channel};
\draw [thick, ->] (6.2,0.5)--(7,0.5);
\node[draw =none] at (6.6,0.65) {$Y_i$};
\draw (7,0) rectangle (8.4,1);
\node[draw =none] at (7.7,0.5) {$\phi(Y^n)$};
\draw [thick, ->] (8.4,0.5)--(9,0.5);
\node[draw =none] at (8.7,0.75) {$\hat M$};
\draw [thick, ->] (5.6,-0.5)--(5.6,0);
\node[draw =none] at (5.6,-0.7) {$U^n$};
\draw [thick, ->] (5.6,1)--(5.6,2.5)--(4,2.5);
\draw (1,2) rectangle (4,3);
\node[draw =none] at (2.5,2.5) {$g(X^{i-1}, Z^{i-1})$};

\draw [thick, ->] (1,2.5)--(0,2.5);
\node[draw =none] at (0.5,2.75) {$\hat S_i$};
\node[draw =none] at (6,1.2) {$Z_{i-1}$};
\draw [thick, ->] (5.6,1.2)--(3,1.2)--(3,1);
\draw [thick, ->] (4.55,0.5)--(4.55,2.25)--(4,2.25);
\draw [dashed] (0.8,-0.5) rectangle (4.1,3.5);
\node[draw =none] at (2.5,-0.8) {Transmitter};
\draw[thick, ->] (2.5,2)--(2.5,1);
\end{tikzpicture}
\vspace{-0.5cm}
\caption{system model}
\label{fig1}
\vspace{-0.5cm}
  \end{figure}

\section{Problem Setup} \label{sec:setup}

Consider the point-to-point setup in Figure~\ref{fig1} where a dual function ISAC transmitter (Tx) wishes to communicate the message $M \in [1:2^{n\R}]$ to a receiver (Rx), where $\R$ denotes the rate of communication and $n$ the blocklength of transmission.  At the same time, the Tx also collects backscattering signals to estimate sensing parameters of the system. 

More specifically, at each discrete-time   $i \in \{1, \ldots, n\}$, the Tx emits a channel input to the system, and the environment creates two outputs: the receive signal $Y_i$ observed at the Rx and the backscattered signal $Z_i$ observed at the Tx. Both signals depend on an internal state sequence $U_1, \ldots, U_n$ of the environment. In practical contexts, this state sequence  can model obstacles, fading phenomena, or also velocities or directions of vehicles in the neighborhood of the Tx or the Rx. We formalize a very general model with channel transition law
\begin{equation}
P_{Z_iY_i| X^{i} U^i Z^{i-1} Y^{i-1}}, \quad i \in\{1,\ldots, n\}, 
\end{equation}
in function of the current and past inputs and states and the past outputs and backscatterers. 
Both the channel from the Tx to the Rx (also called communication channel) as well as the channel from the Tx to itself (also called sensing or radar channel) can thus have arbitrary memory and arbitrarily depend on the past  state symbols $U^i$. The two channels can either be dependent or independent, depending on the specific choice of channel laws $\{P_{Z_iY_i| X^{i} U^iZ^{i-1} Y^{i-1}}\}$ that one considers. 

The Tx can compute its channel inputs in an interactive way, depending on the previously observed backscatterers. Thus, at a given time $i \in \{1, \ldots, n\}$, the Tx produces the input $X_i$ as 
\begin{equation}
X_i = f_i(M, Z^{i-1}, X^{i-1})
\end{equation}
using some encoding function $f_i$ on appropriate domains. 

In terms of sensing, the Tx is interested in estimating a given target $S_i$ which can depend on the channel's internal state $U_i$ and inputs/outputs.  The Tx produces a time-$i$ state  estimate $\hat{S}_i$ in an online manner based on its previous observations: 
\begin{equation}
\hat S_i= g_i(X^{i}, Z^{i}).
\end{equation}
Sensing performance is measured by average block distortion: 
\begin{IEEEeqnarray}{rCl}
\Delta^{(n)} : = \frac{1}{n} \sum_{i = 1}^n \mathbb E [d(S_i, \hat S_i)],
\end{IEEEeqnarray}  
 where  $d(\cdot, \cdot)$  is a given bounded per-symbol distortion function. 
 
The  optimal estimator $g_i^*(\cdot, \cdot)$ is easily obtained from the target distribution $P_{S_i|X^{i},Z^{i}}$ implicitly defined by the channel. We wish to set: 
  \begin{IEEEeqnarray}{rCl} \label{eq:ustar}
\hat{S}_i &: =& \underset{\hat s}{\mathrm{argmin}} \sum_{s}  P_{S_i|X^{i}Z^{i}}(s_i|x^{i},z^{i}) d(s_i, \hat s).
  \end{IEEEeqnarray}

The receiver waits until it observes all its $n$ channel outputs $Y^n$ and then  decodes message $M$ as
\begin{IEEEeqnarray}{rCl}
\hat M &=& \psi(Y^n),
\end{IEEEeqnarray} 
using a well-chosen decoding function $\phi$ that acts on appropriate domains. (The Rx thus does not have explicit knowledge of the state \`a priori, only what it learns from its observed outputs.) 

In an information-theoretic tradition, communication performance is measured by the rate $\R$ that allows to drive the Rx's error probability  $\epsilon^{(n)} : = \Pr [ \hat M \neq M]$ to $0$ asymptotically. This is formalized in the following section.

\section{Capacity-Distortion Tradeoff}

\begin{definition}
A rate-distortion pair $(\R, \D)$ is said to be achievable if there exists a sequence (in $n$) of $(2^{n\R},n)$ codes and encoding, estimation and decoding  functions $f_1, \ldots, f_n, g_1, \ldots. g_n, \psi$ that simultaneously satisfy 
\begin{IEEEeqnarray}{rCl}
\lim_{n \to \infty} \epsilon^{(n)} &=& 0, \\
\lim_{n \to \infty} \Delta^{(n)} & \le & \D.
\end{IEEEeqnarray}
The capacity-distortion trade-off $\mathsf C (\D)$ is the largest rate $\R$ such that the rate-distortion tuple $(\R, \D)$ is achievable.
\end{definition}

The capacity-distortion  trade-off $\mathsf C (\D)$ of our model can be obtained  following similar steps to \cite{Chen2023}, see the following Proposition~\ref{prop:1}. The only difference between the model here and the one in \cite{Chen2023} lies in the way the transmitter  estimates the state sequence. While in \cite{Chen2023}, the state estimation is performed \emph{at the very end} of the communication, here we impose \emph{online estimators} where the $i$-th state symbol has to be estimated at the same time as producing the $i$-th channel inputs. 
\begin{proposition}\label{prop:1}
The  capacity-distortion trade-off $\mathsf C (\D)$ is 
\begin{IEEEeqnarray}{rCl} \label{eq:C}
\mathsf C(\D) = \lim_{n \to \infty} \frac{1}{n}\;  \sup_{\{P_{X_i|X^{i-1} Z^{i-1}} \}_{i=1}^n}\;  \sum_{i = 1}^n I(X^i; Y_i | Z^{i-1}), \IEEEeqnarraynumspace \\
\text{subject to } \; \;  \frac{1}{n} \sum_{i = 1}^n \mathbb E [ d(S_i, g_i^*(X^{i}, Z^{i})) ] \le \D,\IEEEeqnarraynumspace
\end{IEEEeqnarray}
where $g_i^*(\cdot, \cdot)$ is the argmin-estimator in \eqref{eq:ustar}. 
\end{proposition}
Above formula for the capacity-distortion trade-off is difficult to evaluate due to the limit $n\to \infty$  and the supremum over the conditional laws. It simplifies for certain classes of channels, such as obviously memoryless channels or the set of \emph{unifilar channels} \cite{Gallager1968} on which we shall focus in this  article.

\begin{definition} \label{def3}
Consider perfect feedback, i.e., $Y=Z$. A state-dependent channel is called \emph{unifilar} if 
\begin{subequations}
\begin{IEEEeqnarray}{rCl}
P(y_i| x^i, y^{i-1}, u^{i}) =  P(y_i| x_i, u_i) \IEEEeqnarraynumspace
\end{IEEEeqnarray}
and 
\begin{equation}
u_i = \phi(x_i, u_{i-1}, y_i)
\end{equation}
\end{subequations}
for a given state-transition function $\phi(\cdot)$ on appropriate domains. 
\end{definition}

\begin{theorem}\label{th1}
 Given $\D$ the  capacity-distortion trade-off of a connected unifilar channel, where the initial state $s_0$ is available to both the encoder and decoder, is given by the following optimization problem:
 \begin{IEEEeqnarray}{rCl} 
 &&\C(\D) = \lim_{n \to \infty} \max_{\{P_{X_i|U_{i-1}Y^{i-1}}\}_{i = 1}^n} \frac{1}{n}\sum_{i = 1}^n I(X_i, U_{i-1};Y_i|Y^{i-1})\IEEEeqnarraynumspace \label{eq:13}\\
\lefteqn{\textnormal{subject to}} \nonumber  \\&& \lim_{n \to \infty} \frac{1}{n} \sum_{i = 1}^n \sum_{y^{i},x^{i}} \hspace{-0cm} \min_{\hat s} \sum_{s}  P_{S_i|X^{i}Y^{i}}(s_i|x^{i},y^{i}) d(s_i, \hat s)  \le \D. \label{eq:14n}  \IEEEeqnarraynumspace
 \end{IEEEeqnarray}
 \end{theorem}
 \begin{IEEEproof}
 The term in \eqref{eq:13} is equivalent to the capacity of a unifilar channel and has been proved in \cite[Theorem~1]{Permuter2008}. The condition in \eqref{eq:ustar} stems from the optimal estimator  in \eqref{eq:ustar}. 
 \end{IEEEproof}

\section{Reinforcement Learning Approach to ISAC} \label{sec:RL}
To evaluate the capacity-distortion trade-off of the proposed ISAC system,  we require to solve a complex multi-letter optimization problem, see Theorem~\ref{th1}. Our approach is to   first present a Markov decision process (MDP)  formulation of  this optimization problem. We then employ RL where we model  the Tx with an agent performing both sensing and communication tasks.  For this purpose, we use the DDPG algorithm \cite{Lillicrap2015} which is an  actor-critic model-free  RL algorithm that operates over continuous action spaces and is of deterministic gradient policy.

\subsection{MDP formulation of the capacity-distortion trade-off}
To formulate the capacity-distortion trade-off as an MDP, we require to determine the triple $(\delta_{i-1}, a_i, r_i)$ where $\delta_{i-1}$ is the state at time $i$, $a_i$ is the action at time $i$ and $r_i$ is the corresponding reward at time $i$.  For some fixed $\beta \in [0, 1]$,  we determine this triple as in Table~\ref{t1}. 
 A new state $\delta_i $ is then generated according to Equation \eqref{eq:14} shown on the next page. 
 \begin{figure*}[b!] 
\hrule
 \begin{IEEEeqnarray}{rCl} \label{eq:14}
 \delta_i(u_i) = \frac{\sum_{x_i, u_{i-1}} \delta_{i-1} (u_{i-1}) a_i (x_i, u_{i-1}) P(y_i|x_i, u_{i-1}) \mathbbm 1 \{ u_i = f(x_i, u_{i-1}, y_i\}}{\sum_{x_i, u_{i-1}, \tilde u_i}  \delta_{i-1} (u_{i-1} )a_i (x_i, u_{i-1}) P(y_i|x_i, u_{i-1}) \mathbbm 1 \{ \tilde u_i = f(x_i, u_{i-1}, y_i\}}.
 \end{IEEEeqnarray}\end{figure*}
 
 \begin{table} 
 \caption{MDP formulation of the capacity-distortion trade-off}
 \begin{center}
 \normalsize
\begin{tabular}{ |c|c| } 
 \hline
 state $\delta_{i-1}$& $P_{U_{i-1}|Y^{i-1}}(\cdot|y^{i-1})$ \\ 
 \hline
 action  $a_i$ & $P_{X_i|U_{i-1} Y^{i-1}} (\cdot|\cdot, y^{i-1})$  \\ 
 \hline
 reward $r_i$ & $I(X_i, U_{i-1};Y_i|Y^{i-1}) - \beta \mathbb E[ d(S_i, \hat S_i)]$.\\ 
 \hline
  disturbance & $y_i$ \\ 
 \hline
\end{tabular}
 \label{t1}
\end{center}
\vspace{-0.1cm}
\end{table}

\subsection{DDPG Algorithm} \label{sec:DDPG}
The training procedure of this algorithm consists of $K$ episodes each of $T$ sequential steps \cite{Lillicrap2015}. In each step, we perform the following two operations: 

\subsubsection{Collecting Experience at the Agent}
Given the current state $\delta_{i-1}$, the agent takes an action $a_{i} = A_{\mu} (\delta_{i-1})$ according to the $\epsilon$-greedy policy as follows:
\begin{IEEEeqnarray}{rCl}
a_{i} &= &\begin{cases} \underset{\tilde a_1}{\mathrm{argmax}} \; Q_{\pi} (\delta_{i-1},\tilde a_1), & \text{w.p.} \; \; 1- \epsilon, \\
\text{A random action}  &  \text{w.p.} \; \;  \epsilon,
\end{cases} \IEEEeqnarraynumspace
\end{IEEEeqnarray}   
with $\epsilon \in [0, 1]$. 
   After taking the action $a_{i}$, the agent observes the incurred reward $r_i$ and the new state $s_i$. 
The tuple $\{ \delta_{i-1}, a_{i},  r_i, \delta_i\}$ is then stored in a \emph{replay buffer} denoted by $\mathsf B$. 

\subsubsection{ Improving agents and environment networks}
 Draw $N$ samples randomly from $\mathsf B$. For each transition $j \in \{1, \ldots, N\}$, we compute the sampled estimate for future rewards denoted by $b_j$. We then minimize the following objective function: 
\begin{equation} \label{eq:loss}
L(w) = \frac{1}{N} \sum_{j = 1}^N \left ( Q_{w} ( \delta_{j-1}, A_{\mu} (\delta_{j-1})) - b_j\right)^2
\end{equation}
over the parameters of the environment network $w$. 
To maximize the estimate of future cumulative rewards, the parameter  $\mu$ is updated as follows 
\begin{IEEEeqnarray}{rCl}
\mu&\to & \mu+ \frac{\eta}{N} \sum_{j = 1}^N \grad_{A} Q_w\left (\delta_{j-1}, A \right) |_{A= A_{\mu} (\delta_{j-1})} \grad_{\mu} A_{\mu} (\delta_{j-1}), \nonumber\\
\end{IEEEeqnarray}
where $\eta$ is the learning rate at the agent.

%
%

\section{Example: Binary Channel with Multiplicative Bernoulli State}
Consider the channel 
\begin{equation} \label{eq:channel}
Y_i = S_iX_i, \quad i \in \{1, \ldots, n\}
\end{equation}
with binary alphabets $\mathcal X = \mathcal S = \mathcal Y \in \{0,1\}$. Assume that  the feedback is perfect, i.e., $Z_i = Y_i$, and that the target sequence satisfies \begin{IEEEeqnarray}{rCl}
S_i = S_{i-1} \oplus \tilde S_i , \qquad i=1, 2, \ldots,\label{eq:state_evol}
 \end{IEEEeqnarray}
 with  $\{\tilde S_i\}$ being i.i.d Bernoulli$(p)$ and  $S_0 = 0$ deterministically. We consider the Hamming distortion measure $d(s, \hat s) = s \oplus \hat s$.  
 
For each time $i=1,2, \ldots$, and depending on the past inputs $X_1, \ldots, X_{i-1}$,  let $L_i$ denote the time evolved  since the input was $1$ for the last  time. This is, $L_i$ is such that 
\begin{equation}
X_{i-1}=\cdots=X_{i-L_{i}+1}=0 \quad \textnormal{ and } \quad X_{i-L_i}=1.
\end{equation} Then, define the time-$i$ auxiliary random variable  \begin{equation}\label{eq:si}
 U_{i} := (L_i, S_{i-L_i}).
 \end{equation} 
 Notice that 
\begin{eqnarray}
U_i &=& \phi( X_i, Y_i, U_{i-1}) =\phi( X_i, Y_i,(L_{i-1}, S_{i-1-L_{i-1}})) \nonumber \\&=& \begin{cases}
( L_{i-1}+1, S_{i-L_{i-1}}) & \textnormal{ if } X_i=0  \\
(0, Y_i /X_i ) &  \textnormal{ else}. 
\end{cases} \label{eq:stat}
\end{eqnarray}  
In other words, the Tx can calculate the auxiliary sequence $\{U_i\}$ using an online procedure. 

We next determine the optimal estimator to estimate the target $S_i$ based on $X^i$ and $Y^i$. When $X_i=1$, the Tx should obviously set $\hat{S}_i=Y_i$ resulting in distortion $d(S_i, \hat{S}_i)=0$.  To understand how to estimate $S_i$ when $X_i=0$, notice that by \eqref{eq:state_evol}:\begin{equation}
S_i=S_{i-L_{i}} \oplus \tilde S_{i-L_{i}+1}  \oplus \cdots \oplus  \tilde S_{i} . 
\end{equation}
Conditioned on $S_{i-L_i}$, it is thus trivially independent of inputs, outputs, and states prior to time $i-L_{i}+1$. Moreover,  since by definition of $L_{i}$ we have $X_{i-L_i+1}=\cdots =X_{i-1}=0$, it is also independent of inputs and outputs after time $i-L_i +1$. The optimal estimator is thus the maximum likelihood estimator based on $U_{i}=(L_i, S_{i-L_i})$, which sets
\begin{equation}
\hat{S}_{i} = S_{i-L_i} \oplus \mathbbm{1}\{p_{L_i}> 1 - p_{L_i}\} 
\end{equation}
where we define (notice that the right-hand side in the following expression does not depend on $\ell$):
\begin{equation}\label{eq:defpl}
p_{\ell} : = \Pr[ \tilde S_{i-\ell+1}  \oplus \cdots \oplus  \tilde S_{i}=1]. 
\end{equation} 
Conditioned on the value of $L_i=\ell$ and given that $X_i=0$, the    distortion for the time-$i$ symbol is thus 
\begin{equation}\label{eq:dist}
d(S_i, \hat{S}_i)=\min\{ p_{\ell}, 1 - p_{\ell}\}. 
\end{equation}

Following a similar reasoning as in the derivation of the distortion under $X_i=0$, we can conclude that  given $(U_i, X_i)$ the channel output and the feedback signal are conditionally independent of the previous inputs, outputs, and auxiliaries: 
\begin{equation} \label{eq:MC}
(X^{i-1}, Y^{i-1}, U^{i-1}) \to (X_i, U_i) \to Y_i, \quad i=1,2,\ldots
\end{equation}
Combined with \eqref{eq:stat},  this establishes  that channel in this example can be seen as 
a unifilar channel according to Definition~\ref{def3}, where the auxiliaries $U_i$ acts as the time-$i$ state and  the 
channel transition law  is:
\begin{equation*}
\Pr[ Y_i = y | X_{i}=x, U_{i}=(\ell, s)] = \begin{cases} 1 & y=x=0 \\
0 & y=1, x=0 \\
p_\ell & y\neq s,  x=1\\
1- p_\ell & y=s, x=1. 
\end{cases}
\end{equation*}
Notice that by solving the linear recursion $p_\ell= (1-p)p_{\ell-1}+ p (1-p_\ell)$, one obtains that 
 \begin{equation}\label{eq:pli}
 p_{\ell} =\frac{1}{2} - \frac{1}{2} (1-2p)^{\ell}.
 \end{equation}

We evaluate  Theorem~\ref{th1} for our unifilar channel. 
 
 \begin{theorem}\label{th2}
Given $\D$, the capacity-distortion trade-off of the binary channel with multiplicative state in \eqref{eq:channel} is given by the following optimization problem:
\begin{subequations}\label{eq:subCD}
\begin{IEEEeqnarray}{rCl} 
\C(\D) &&= \lim_{n \to \infty} \max_{\mathcal P} \frac{1}{n}\sum_{i = 1}^n  H_b\left (\sum_{\ell = 1}^{i-1} \alpha_i (\ell) \right ) - \sum_{\ell = 1}^{i-1} \kappa_i(\ell) H_b(p_{\ell}), \nonumber\\  \\
\text{s.t.:} &&\;  \sum_{i = 1}^n \sum_{ \ell = 1}^{i-1}  \Pr [X_i = 0, \ell_i = \ell ]\min\{p_{\ell}, 1-p_{\ell} \}\le n \D. 
\end{IEEEeqnarray}
\end{subequations}
where $H_b(\cdot)$ is the binary entropy function and
\begin{IEEEeqnarray}{rCl}
\mathcal P & : = & \{P_{X_i|U_{i-1}}(x_i|u_{i-1})\}_{i = 1}^n,\\
\kappa_i (\ell)& : =&  \Pr [X_i = 1,  L_{i-1}= \ell ], \label{eq:kappai} \\
\alpha_i (\ell)& :=&   \Pr [X_i = 0, L_{i-1} = \ell]  \notag \\
&&+  \Pr [X_i = 1, L_{i-1} = \ell, S_{i-L_i} = 1 ]\cdot p_{\ell}  \notag \\
&&  + \Pr [X_i = 1,L_{i-1} = \ell, S_{i-L_i} = 0 ] \cdot(1-p_\ell). \label{eq:alphai}
\end{IEEEeqnarray}
\end{theorem}
\begin{IEEEproof}
We have for the $i$-th term: 
\begin{IEEEeqnarray}{rCl}
\lefteqn{I(X_i, U_{i-1}; Y_i|Y^{i-1})} \notag \\
 &=& H(Y_i | Y^{i-1}) - H(Y_i|X_i, U_{i-1}, Y^{i-1}) \\
& = & H_b(\Pr [Y_i = 0| Y^{i-1} = y^{i-1}]) \notag \\
&&-  \sum_{\ell = 1}^{i-1} \Pr [X_i = 1, L_i=\ell] H_b (p_{\ell}) \IEEEeqnarraynumspace\\
 & = & H_b\left( \sum_{\ell = 1}^{i-1} \alpha_i(\ell)\right) -  \sum_{\ell = 1}^{i-1} \kappa_i(\ell) H_b(p_{\ell}), \label{eq:comm2}
\end{IEEEeqnarray}
where $\kappa_i$ and $\alpha_i$ are defined in \eqref{eq:kappai} and \eqref{eq:alphai}, respectively. Note that the first term in \eqref{eq:comm2} is due to the fact that 
\begin{IEEEeqnarray}{rCl}
\lefteqn{\Pr[ Y_i = 0 | Y^{i-1} = y^{i-1}] } \notag \\
&=& \sum_{\ell = 1}^{i-1} \sum_{s} \sum_x \Pr [X_i = x, U_{i-1} = (\ell, s) ] \notag \\
&& \hspace{0.7cm}\Pr[ Y_i = 0 | Y^{i-1} = y^{i-1}, X_i = x, U_{i-1} = (\ell, s)] \\
&=& \sum_{\ell = 1}^{i-1}  \Pr [X_i = 0, L_i = \ell] \notag \\
&+&   \sum_{\ell = 1}^{i-1}  p_{\ell}  \cdot \Pr[ X_i = 1, L_i = \ell, S_{i-L_i} = 1 ] \\
&+&    \sum_{\ell = 1}^{i-1} (1- p_{\ell}) \cdot  \Pr[ X_i = 1, L_i = \ell, S_{i-L_i} = 0 ].
\end{IEEEeqnarray}
The second term follows by the fact that conditioning on $L_i = \ell$,  if $X_i = 0$ then $H(Y_i |X_i, U_{i-1}, Y^{i-1}) = 0$, and if $X_i = 1$ then  $H(Y_i |X_i, U_{i-1}, Y^{i-1}) = H_b(p_{\ell})$.

The distortion calculation follows immediately from \eqref{eq:dist} and because when $X_i=1$ then the $i$-th distortion term is zero.
\end{IEEEproof}

\section{Numerical Analysis}
In our implementation, we  trained the agent for $500$ episodes, each containing $100$ consecutive blocks. The Monte Carlo evaluation  length for the average reward was chosen to be $1000$. We solve the optimization  problem in \eqref{eq:subCD} using an RL approach with full (unbounded), limited  and degenerate sate spaces. In the limited case, we restrict the algorithm state space to a fraction $k \in [0,1]$ of the full state space $\delta_{i-1}$. In the degenerate case, the optimization problem in \eqref{eq:subCD} is solved for memoryless strategies, i.e., for $X_i$ independent of $U _{i-1}$, or equivalently using an RL approach with constant state-space.

\begin{figure}[t]
\center
				\includegraphics[width=0.37\textwidth]{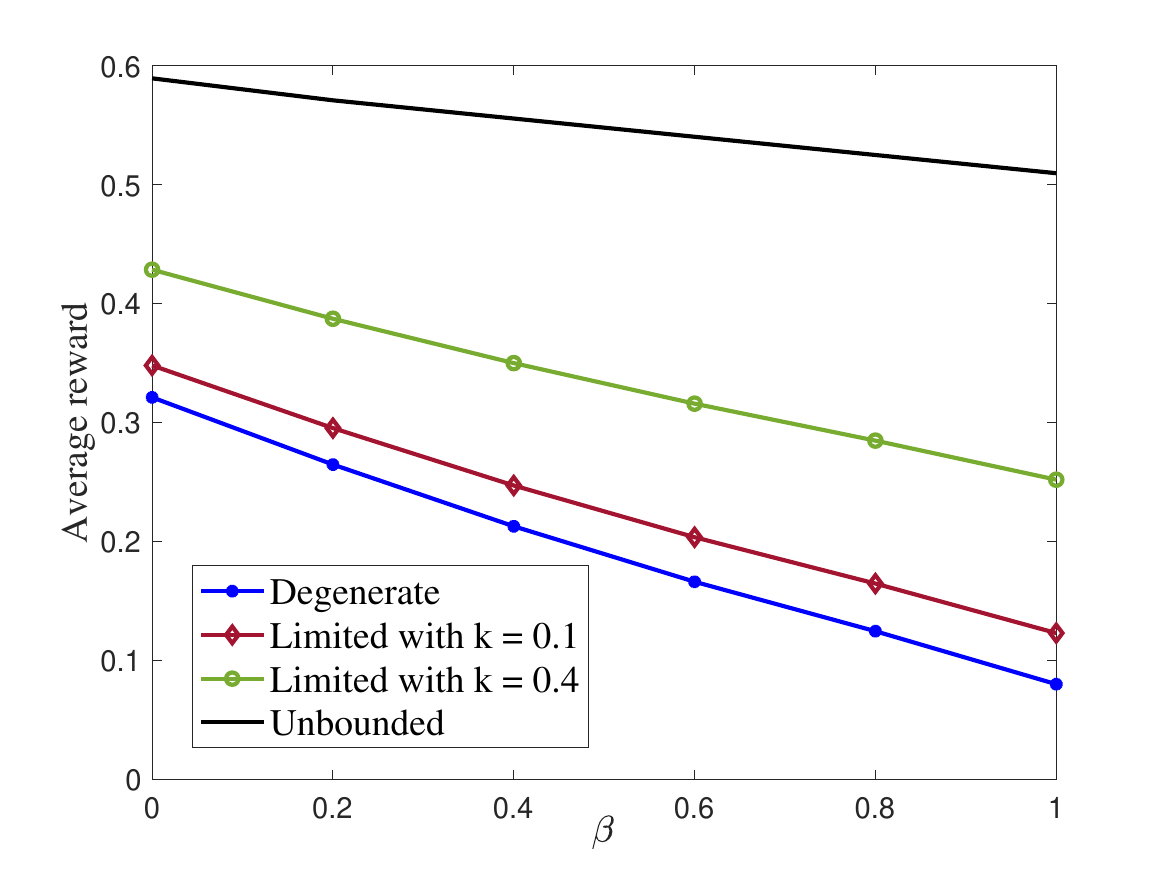}
\caption{Average reward as a function of $\beta$ with $n = 5000$ and $p = 0.1$ for degenerate, limited with $k \in \{0.1,0.4\}$ and unbounded state spaces.}
\label{fig2}
\vspace{-0.5cm}
\end{figure}
\begin{figure}[t]
\center
				\includegraphics[width=0.45\textwidth]{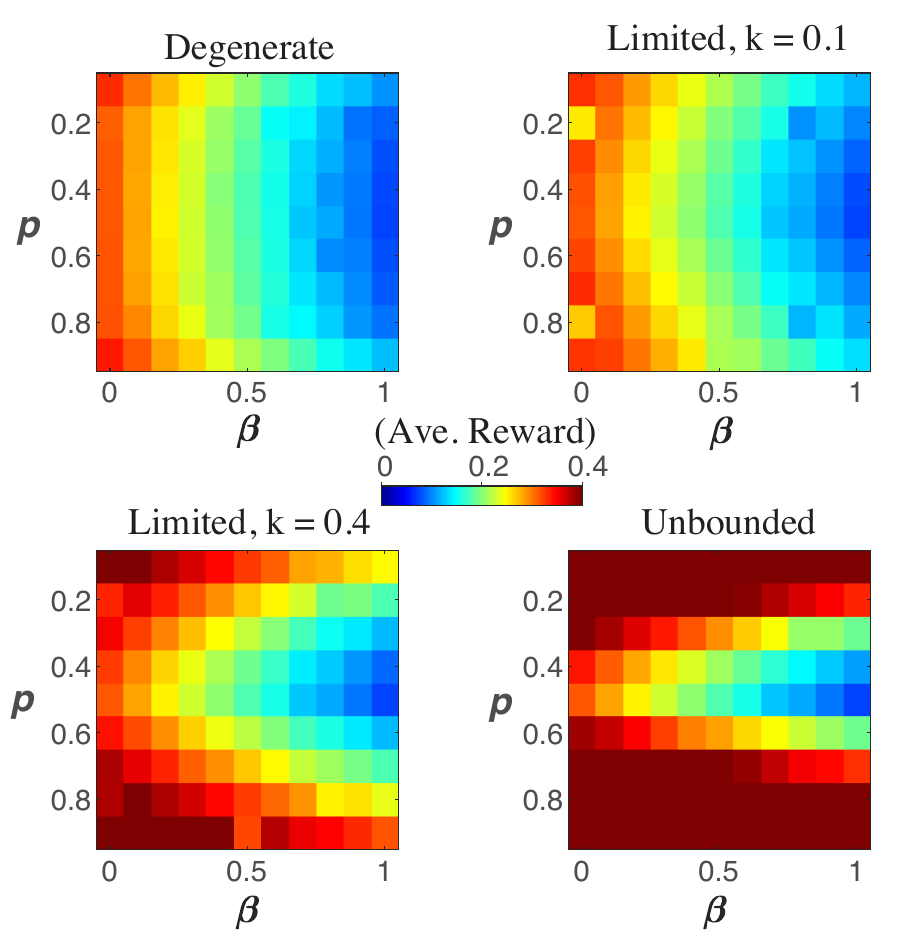}
\caption{Average reward as a function of $p$ and $\beta$ for  $n = 5000$ for degenerate, limited with $k \in \{0.1,0.4\}$ and unbounded state spaces.}
\label{fig5}
\vspace{-0.5cm}
\end{figure}
			
			\begin{figure}[t]
\center
				\includegraphics[width=0.37\textwidth]{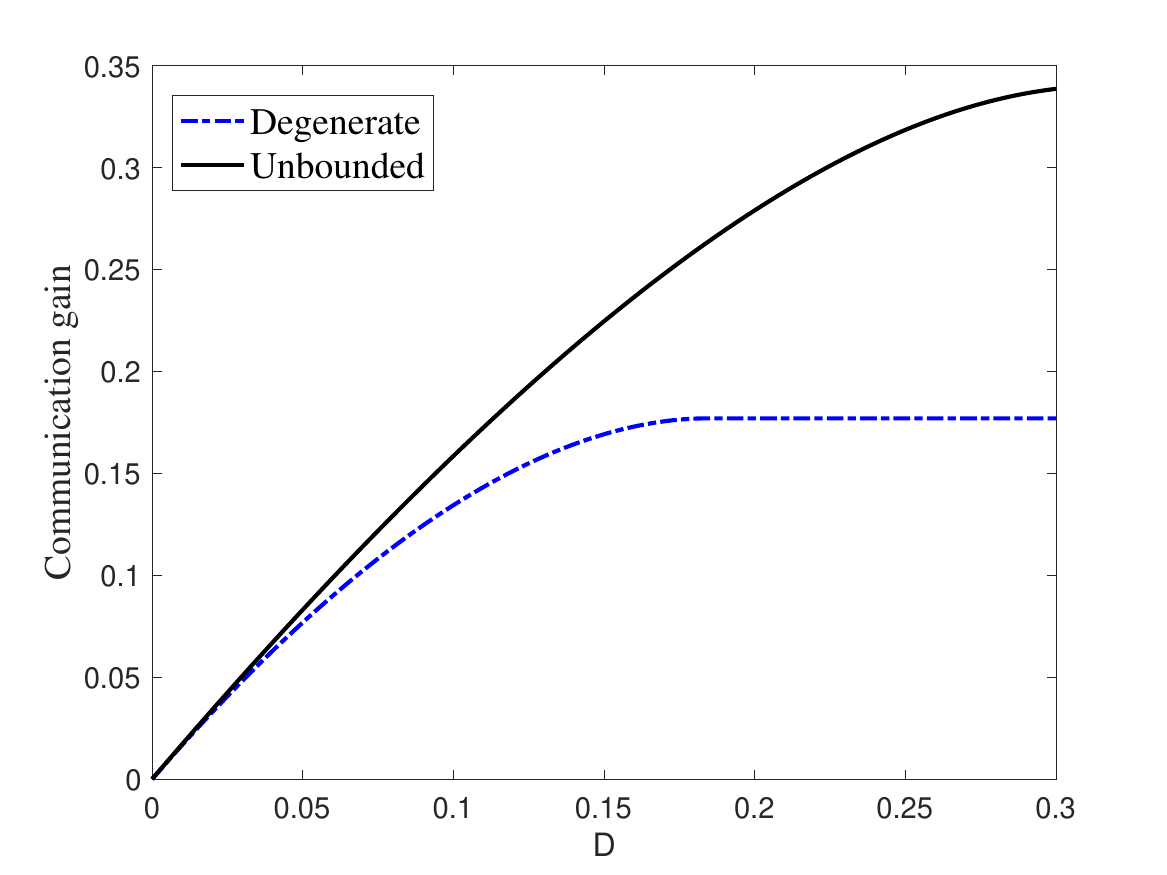}
\caption{Communication gain versus  the maximum distortion $\D$ with $p = 0.3$ for degenerate and unbounded state space cases.  }
\label{fig3}
\vspace{-0.5cm}
\end{figure}				
Fig.~\ref{fig2} illustrates the average reward versus the parameter $\beta$ for  $n = 5000$ and $p = 0.1$ for unbounded, limited with $k \in \{0.1, 0.4\}$ and degenerate state spaces. In a similar way, 
 Fig.~\ref{fig5} illustrates this average reward as a function of both $\beta$ and $p$. We observe that  the average reward is high for very large ( i.e., $p> 0.8$) and very small (i.e., $p<0.2$) values of $p$, and generally whenever $p\neq 0.5$, the average reward is strictly larger with unbounded state space  compared to the limited and degenerate cases.   For $p = 0.5$, the  situation is somehow degenerate and average reward is the same for the cases with degenerate, limited and unbounded state spaces. 
 
 Fig.~\ref{fig3} illustrates the communication gain versus $\D$ for the cases with unbounded and degenerate state spaces when $p = 0.3$.  As can be seen from this figure, enlarging the state space in our RL framework significantly improves sensing and communication performances. 
  
  \section{Conclusions}
We have considered an ISAC system where  a transmitter sends a message to a receiver over a channel with memory and simultaneously estimates given targets by analyzing the backscattered signals from the emitted waveform. Estimation of the targets was performed in an online matter. We have used Massey’s concept of directed information to derive the capacity-distortion trade-off for this ISAC setup  and simplified the expression for the class of unifilar channels. We then presented an MDP formulation of the resulting waveform optimization problem and solved it by employing the DDPG algorithm.  Our numerical results have shown a significant performance improvement when the RL approach can take advantage of the full (unbounded) state space as compared to models with limited (or even degenerate) state spaces. 
\section*{Aknowledgment}
The work of H. V. Poor has been supported by the  U.S National Science Foundation under Grant ECCS-2335876. The work of S. Shamai was supported by the German Research Foundation (DFG) via the German-Israeli Project Cooperation (DIP), under Project SH 1937/1-1, and by the Ollendorff Minerva Center of the Technion. The work of M. Wigger has been supported by the European Research Council under Grant Agreement 101125691. 

\appendices 

\end{document}